\documentclass[epsfig,12pt]{article}
\usepackage{epsfig}
\usepackage{graphicx}
\usepackage{array}
\usepackage{color}
\usepackage{bm}
\usepackage{cite}
\usepackage{geometry}
\usepackage{latexsym}
 \usepackage{amsmath, amssymb, amscd, xypic, graphicx}

\newcommand{\beq}{\begin{equation}}   
\newcommand{\eeq}{\end{equation}}
\newcommand{\beqn}{\begin{eqnarray}}   
\newcommand{\eeqn}{\end{eqnarray}}

\def\tr{\text{tr}}
\def\N{{\mathcal N}}

\def\Z{{\mathbb Z}}

\newcommand*\xbar[1]{%
 \kern0.5ex%
  \hbox{%
   \kern0.2ex%
      \vbox{%
      \hrule height 0.5pt % The actual bar
      \kern0.5ex%         % Distance between bar and symbol
      \hbox{%
        \kern-0.1em%      % Shortening on the left side
        \ensuremath{#1}%
        \kern-0.1em%      % Shortening on the right side
      }%
    }%
  }%
}

\newcommand{\gsim}{\lower.7ex\hbox{$
\;\stackrel{\textstyle>}{\sim}\;$}}
\newcommand{\lsim}{\lower.7ex\hbox{$
\;\stackrel{\textstyle<}{\sim}\;$}}
\begin{document}

\begin{titlepage}

\begin{flushright}
FTPI-MINN-15-07, UMN-TH-3421-15,  \\
February 23, 2015
\end{flushright}

\vspace{1cm}

\begin{center}
{  \Large \bf IR  Renormalons vs. Operator Product Expansion  \\[2mm]
in Supersymmetric and Related Gauge Theories
}

\end{center}

\begin{center}
{\large
Gerald V. Dunne,$^{a}$ M. Shifman,$^{b}$  and Mithat \"Unsal$^{\,c}$}
\end {center}

\vspace{1mm}

\begin{center}

$^{a}${\it Department of Physics, University of Connecticut, Storrs,  CT 06269, USA}\\[1mm]
$^b${\it  William I. Fine Theoretical Physics Institute,
University of Minnesota,
Minneapolis, MN 55455, USA}\\[1mm]
$^{c}${\it Department of Physics, North Carolina State University, Raleigh, NC 27695, USA}\\[1mm]

\end{center}

\vspace{0.6cm}

\begin{center}
{\large\bf Abstract}
\end{center}
We use the ``conspiracy" between infrared (IR) renormalons and condensates in the operator product expansion for correlation functions to make predictions concerning the structure of singularities in the Borel plane for the perturbative series in quantum field theories with different levels of supersymmetry. The same conspiracy can be used for establishing the absence of condensates or IR renormalons in gauge theories with an IR conformal regime or fully Higgsed gauge theories. The absence of the renormalon-induced factorial divergence implies that instanton contributions (where present) must be well-defined. We show that the conventional  bubble-chain method for detecting renormalon-induced factorial divergences in these theories is not sufficient.

\end{titlepage}

{\em Introduction}: One of the remarkable successes of the operator product expansion (OPE) is its capacity to connect perturbative information with non-perturbative condensates in a manner that consistently identifies the factorial divergences associated with infrared (IR) renormalons with certain condensates of specific dimensions 
\cite{Wilson:1969zs,Shifman:1978bx,thooft,mueller,beneke,3} (the so-called ``conspiracy''). The classic example is the relation between the first IR renormalon Borel pole in Yang-Mills theory and the OPE gluon condensate,
\begin{eqnarray}
t_{\rm IR}^{(0)}=\frac{32\pi^2}{\beta_0\, g^2(Q^2)} \leftrightarrow \tr\,\langle F^2\rangle \leftrightarrow \left(\frac{\Lambda^2}{Q^2}\right)^2  \,,
\label{one}
\end{eqnarray}
where $\beta_0$ is the first coefficient of the $\beta$ function. The
standard method of identifying renormalons in QCD correlation functions at large momenta is through quark bubble chains, with the subsequent replacement  $-2N_f/3 \to \beta_0$, and the result (\ref{one}) may be interpreted as a confirmation that this replacement works in asymptotically free theories \cite{beneke,3}.

Recently it has been noted in several contexts that this connection is much less straightforward in supersymmetric field theories \cite{Argyres:2012ka,Dunne:2012ae,Shifman:2014fra,Shifman:2014cya}.  In this Letter we investigate renormalons in super-Yang-Mills theories (SYM) and discover novel features. We also consider theories which reduce to SYM in the planar limit, through large-$N$ orbifold/orientifold equivalences \cite{Kovtun:2003hr,Armoni:2003gp,Unsal:2007fb}.  In all these cases we show that the simple bubble-chain graphs, a usual renormalon signature, do not provide meaningful data.

The fact that the correspondence between diagrammatic renormalon arguments and the OPE fundamentally differ  for SUSY theories can be understood from the simple observation that all purely gluonic operators have vanishing condensates in SYM theory without matter \cite{Shifman:2014fra}. Thus, consistency between the OPE and the perturbative renormalon analysis requires that certain renormalon Borel poles must have vanishing residues, which must happen via intricate cancellations between classes of diagrams. These cancellations are not captured by any straightforward diagrammatic arguments.

We turn this observation around, and use the general structure of the OPE as a tool to reveal the behavior of various perturbative expansions. The renormalon-related factorial divergence can appear only if it has a chance of conspiracy. In other words, renormalons can generate only such degree of the factorial perturbative divergence  that can conspire with the {\em nonvanishing} OPE condensates. Since in supersymmetric theories the class of possible condensates is more restricted than in Yang-Mills theory, a  number of conventional renormalons should simply disappear. This approach is also motivated by an attempt to interpret the OPE in terms of the formalism of resurgent trans-series.
In the concluding part we discuss some nonsupersymmetric Yang-Mills models with an IR conformal regime, which also have peculiar renormalon patterns. 

\vspace{1mm}

\begin{figure}[htb]
 \center
  \includegraphics[scale=0.28]{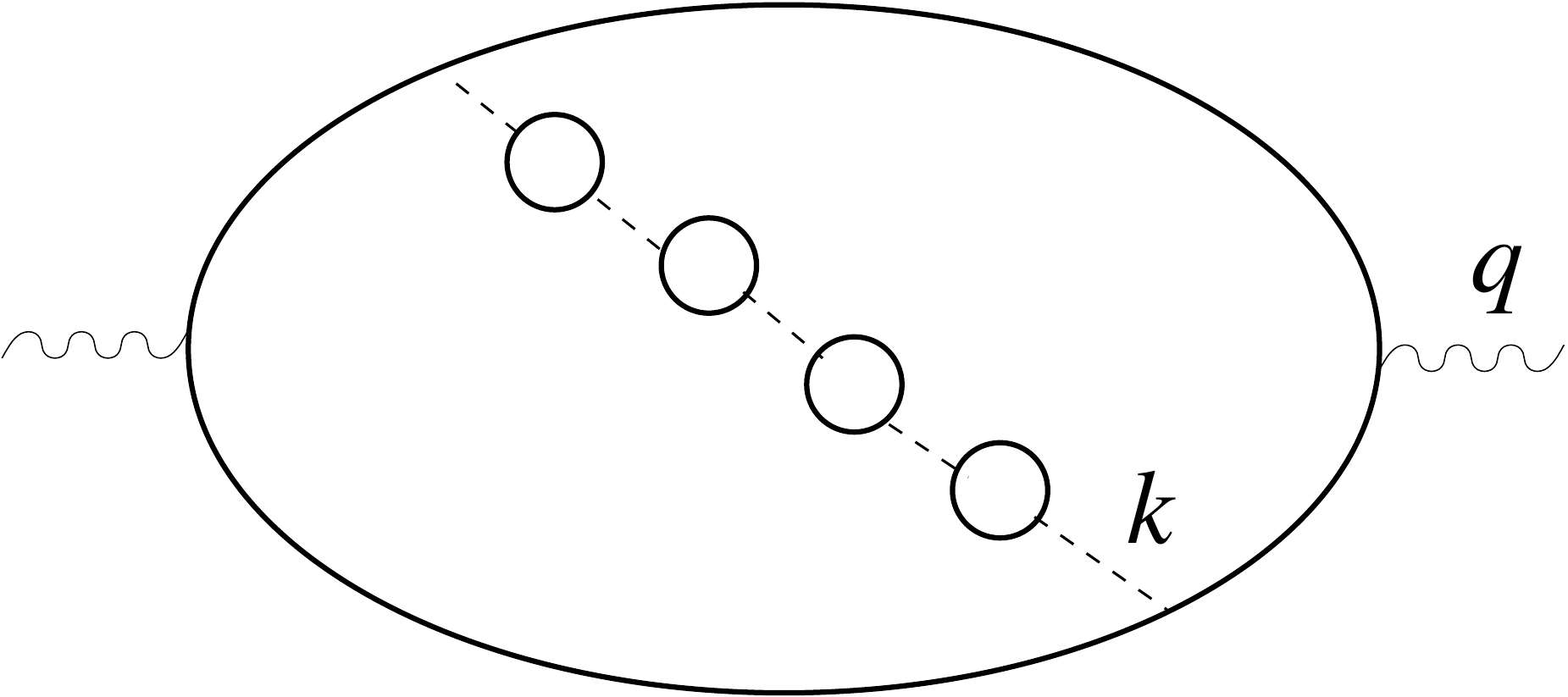}
  \caption{The bubble-chain diagram for a two-point function associated with renormalons. Solid lines denote quark propagators, while dashed lines are for gluons. }
  \label{f1}
  \end{figure}

{\em Bubble graphs as a detector for renormalons}: The standard method of isolating the renormalon 
contributions is through insertions of the fermion bubbles in the gluon propagator (for reviews see \cite{beneke,3,Shifman:2014fra}). 
It is crucial that the fermion bubbles appear only in the gluon propagators, and that $N_f$ is a free parameter
(Fig. \ref{f1}). Only in this case the given method allows one to single out renormalons from particular sets of graphs
(Fig. \ref{f1}).
  
  The graph above refers to a two-point function of the type
  \beq
  i\int d^4x e^{iqx}\, \langle J(x) J(0)\rangle\,.
  \eeq
  A typical example is the calculation of the Adler function.
Each bubble in the chain is a fermion loop.
In pure $\N =1$ SYM theory, fermions are present in the form of gauginos. The gaugino bubbles have no free parameter, and, moreover, appear in a similar way (through supersymmetry) both on the gluon and gluino lines. Hence, in pure SYM theory the bubble insertion method does not work. 
To make it work, one can introduce $N_f$ matter supermultiplets in the fundamental representation (i.e quarks and squarks), replacing pure SYM theory by supersymmetric QCD (SQCD). Then one could consider matter superfield bubble insertions in the gauge superfield propagator. This conceptually straightforward strategy shows that -- although the correspondence between renormalon
singularities in the Borel plane  and OPE is maintained -- the isolation of renormalons from individual Feynman graphs
as in Fig. \ref{f1} becomes problematic. 

\begin{figure}[htb]
 \center
  \includegraphics[scale=0.4]{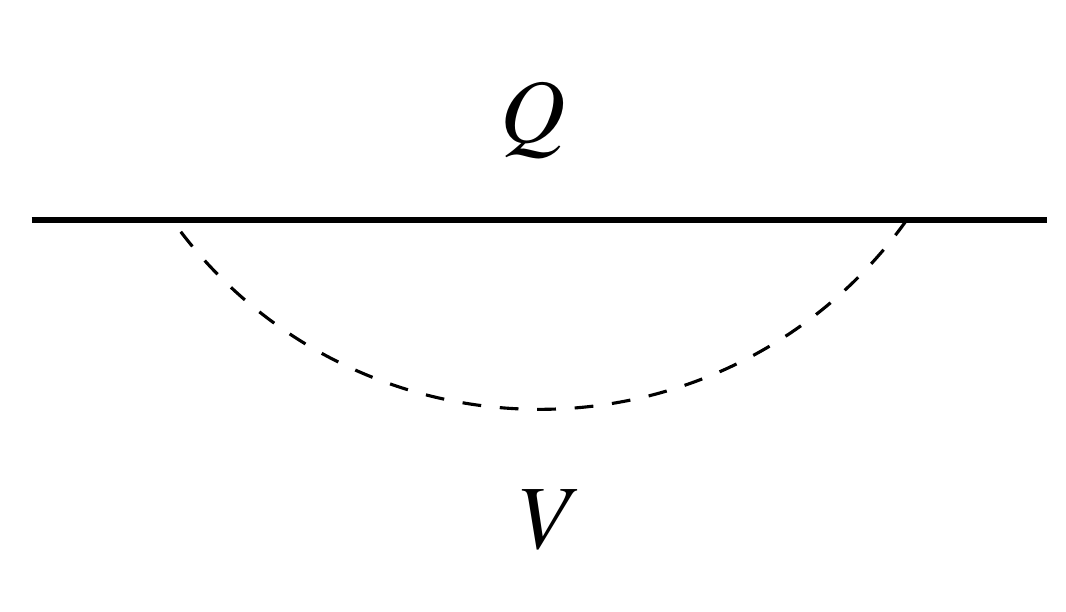}
  \caption{Anomalous dimensions of the matter fields in SQCD. }
\label{figthree}
  \end{figure}

First, let us note that in SQCD the operator basis in the OPE changes as compared to pure SYM. In SQCD the lowest-dimension relevant operator is $\sum_f  \left( \bar Q_f e^V Q^f\right)$ where $Q_f, \tilde Q_f$ are matter superfields. 
Here, the number of matter superfields $N_f$ is assumed to be $N+1 <N_f<3N/2$.
Its lowest component can develop a vacuum expectation value (VEV) which has dimension 2. Hence, instead of (\ref{one}), 
the nearest singularity in the Borel plane is expected to move to
\begin{eqnarray}
t_{\rm IR}^{(0)}=\frac{16\pi^2}{\beta_0\, g^2(Q^2)} \leftrightarrow \sum_f  \left( \bar Q_f e^V Q^f\right) \leftrightarrow\left(\frac{\Lambda^2}{Q^2}\right)  .
\label{two}
\end{eqnarray}
A renormalon signature corresponding to (\ref{two}) can can come only through the
$Z$ factor of the matter field, which is irrelevant in QCD correlation functions. Indeed, the anomalous dimension 
$\gamma$
of the matter field following from the $Z$ factor enters in the two-point
 function of the conserved matter currents as
\cite{Shifman:2014cya}
 \beq D (Q^2)=
  \frac{3}{2}  N \sum_f
q_f^2\left[  1 -   \gamma \left(\alpha_s (Q^2) \right)\right]\,,
\label{m5} 
\eeq 
where $D$ is the Adler function, $f$ is the flavor index, and $q_f$ is the
corresponding charge. The anomalous dimension of the matter fields comes from the diagram depicted in Fig.~\ref{figthree}.
Unlike QCD, the $Z$ factors in SQCD are gauge invariant.
Renormalons corresponding to Fig. \ref{figthree} produce the singularity (\ref{two}) which through (\ref{m5}) penetrates into the Adler function.

\vspace{1mm}

{\em Vanishing VEVs in SYM and suppressed
VEVs in  QCD(AS/S/BF)}: Consider first pure SYM theory. It is well known  that the operator $\tr F^2$, being proportional to the trace of the energy-momentum tensor, cannot develop a nonvanishing VEV.
Vanishing of VEVs of all operators which are built exclusively from the gluon fields 
was proved in \cite{Shifman:2014fra}. The lowest dimension operator
which can appear with a nonvanishing VEV in SYM is $ \langle \tr \lambda \lambda  \bar \lambda \bar \lambda \rangle$
where $\lambda$ is the gluino field.

The lowest component  of    $S= W^{\alpha}W_{\alpha} $ acquiring a  VEV is compatible with unbroken supersymmetry. This is the chiral condensate, 
    \begin{eqnarray}
\big \langle   \textstyle \frac{1}{N}  \tr   \lambda \lambda  \big \rangle = \Lambda^3 e^{\frac{ 2 \pi i k}{N}}, \qquad k=0, \ldots, N-1\,.
\label{chiral}
 \end{eqnarray}
This breaks the $\Z_{2N}$ discrete chiral symmetry down to $\Z_2$, leading to $N$-isolated supersymmetric vacua, $|\Omega_k  \rangle$. 
Since the chiral condensate is charged under the discrete 
chiral symmetry, it cannot appear in the OPE for chiral-invariant correlation functions.
Nor can it
 be associated with a Borel singularity in the Borel plane coming from perturbation theory. Thus, indeed, the first non-vanishing 
 single-trace condensate is
 $\langle \tr \lambda \lambda  \bar \lambda \bar \lambda \rangle$.
This means that in ${\N}=1$ SYM the first ambiguity in the Borel resummation of the perturbation theory must be (at least) of order $\frac{\Lambda^6}{Q^6}$:
\begin{eqnarray}
t_{\rm IR}^{(0)}=\frac{48\pi^2}{\beta_0\, g^2(Q^2)} \leftrightarrow \langle \tr \lambda \lambda  \bar \lambda \bar \lambda \rangle \leftrightarrow \frac{\Lambda^6}{Q^6}  \, .
\label{l6}
\end{eqnarray}
Since IR renormalons are in one-to-one correspondence with the OPE, we conclude that the leading singularity arising from the naive bubble chain diagrams (Fig.~\ref{f1}) must cancel. The gluino bubbles would wrongly identify $\frac{\Lambda^4}{Q^4}$ terms.

Here we will    expand
the class of operators which cannot have condensates in supersymmetric theories. This vanishing severely restricts
 occurrence of renormalons. 

1) For an arbitrary (local) color and Lorentz-singlet composite superfield, only the lowest component 
can have nonvanishing VEVs.

2) If the above operator is a total superderivative, {\em all}  its components must have 
vanishing VEVs. For instance, in SQCD without superpotential, the Konishi anomaly \cite{Kon}
tells us that
\beq
\bar{D}^2 \left( \bar Q_f e^V Q^f\right) = \frac{T(R_f)}{2\pi^2}\, \mbox{tr} \,W^2\,,
\label{8}
\eeq
implying that the gluino condensate (the lowest component of $\mbox{tr} \, W^2$) becomes an order parameter.
It {\em must} vanish if supersymmetry is unbroken. 
One may also use generalized Konishi anomalies \cite{Cachazo} for proving the vanishing of more complicated (higher dimension) condensates.

Now, let us pass to the orbifold/orientifold daughters of 
pure SYM theory, to be referred to as QCD(AS/S/BF) [corresponding to antisymmetric, symmetric and bifundamental fermion representations]. The key observation is that the leading 
renormalons are generated by planar graphs. In the same leading approximation in $1/N$ all expectation values that vanish in SYM and can be projected onto QCD(AS/S/BF) (i.e.  belong to the common sector) will continue 
to vanish in the daughter theories. 
In particular \cite{Armoni:2003gp,Kovtun:2003hr,Unsal:2007fb}, 
\begin{eqnarray}
\big \langle  \textstyle \frac{1}{N} \tr F^2  \big \rangle^{\rm QCD(AS/S) }\sim \frac{1}{N} \Lambda^4 \to 0\,, \quad  
\big \langle  \textstyle \frac{1}{N} (\tr F_1^2 +   \tr F_2^2 ) \big \rangle^{\rm QCD(BF) }\sim \frac{1}{N^2} \Lambda^4
\to 0\,. 
\label{six}
\end{eqnarray}
Combining these two facts together we conclude that the planar renormalons which disappear
in SYM theory must disappear in QCD(AS/S/BF) as well. 

This can also be understood in terms of the connection between the vacuum energy density and the gluon condensate, via the trace anomaly relation in YM,  $\N=1$  SYM,  QCD(AS/S),  and QCD(BF) with massless fermions:
         \begin{align}
{\cal E}_{\rm vac}   =   \frac{1}{4} \langle   T_{\mu \mu}  \rangle    =  \left\{ \begin{array}{ll} 
 - \frac{11 N/3 }{ 64 \pi^2}  \langle \tr F^2  \rangle  & \qquad {\rm YM} \cr \cr
 - \frac{3N }{ 64 \pi^2}  \langle \tr F^2  \rangle  & \qquad {\rm SYM} \cr  \cr
  -\frac{3N +4/3 }{ 64 \pi^2}  \langle \tr F^2  \rangle  & \qquad {\rm QCD(AS)} \cr  \cr
   - \frac{3N -4/3 }{ 64 \pi^2}  \langle \tr F^2  \rangle  & \qquad {\rm QCD(S)}  \cr\cr
   -   \frac{3N }{ 64 \pi^2}  \langle \tr F_1^2 + \tr F_2^2 \rangle  & \qquad {\rm QCD(BF)} \cr  
 \end{array} \right.
 \end{align}
 Note that in the large-$N$ limit the leading order beta-function coincides for all the one-flavor theories, $\N=1$  SYM,  QCD(AS/S),  and QCD(BF).   This is a natural consequence of the orientifold equivalence. 
   Also note that in the sense of the number of  bosonic and fermionic degrees of freedom,  the numbers are almost balanced.  Since ${\rm dim(Adj)} = (N^2-1)  $, there are $(N^2-1)$ gluons in all systems. The number of fermions  are, respectively, 
$
2\, {\rm dim(AS)} = (N^2-N)$, $2\, {\rm dim(AS)} = (N^2+N)$, $2\, {\rm dim(BF)} = (N^2-2)$.
Because of the orientifold/orientifold equivalence, the leading order $O(N^2)$ contribution vanishes in all of these theories.   The vacuum energy starts as  
$ \mu^4 N^1  $ for QCD(AS/S) because the off-set between the number of bosonic and fermionic degrees of freedom is $O(N^1)$. The vacuum energy starts as  
$ \mu^4 N^0  $ for QCD(BF) because the off-set of the number of bosonic and fermionic degrees of freedom is $O(N^0)$.  Thus, from the microscopic point of view, there is an apparent Bose-Fermi cancellation.  This exact cancellation is essentially inherited from $\N=1$ SYM due to the large-$N$ equivalence, despite the fact that  QCD(AS/S/BF) are non-supersymmetric.    The non-perturbative part of the vacuum energy density, by the large-$N$ equivalence,  will be $O(N)$ for QCD(AS/S), and $O(N^0)$ for QCD(BF).

Thus, AS/S/BF theories show that the renormalon factorial divergence generally speaking cannot be detected
through bubble chains. Indeed, in these theories one can replicate fermions, acquiring $N_f$ as a free parameter.
If $N_f\geq 2$ they lose their planar connection with SYM theory, implying occurrence of nonvanishing condensates of the type (\ref{six}).
For conspiracy, so should factorial divergence of the perturbative series.

\vspace{1mm} 

\begin{figure}[htb]
 \center
  \includegraphics[scale=0.4]{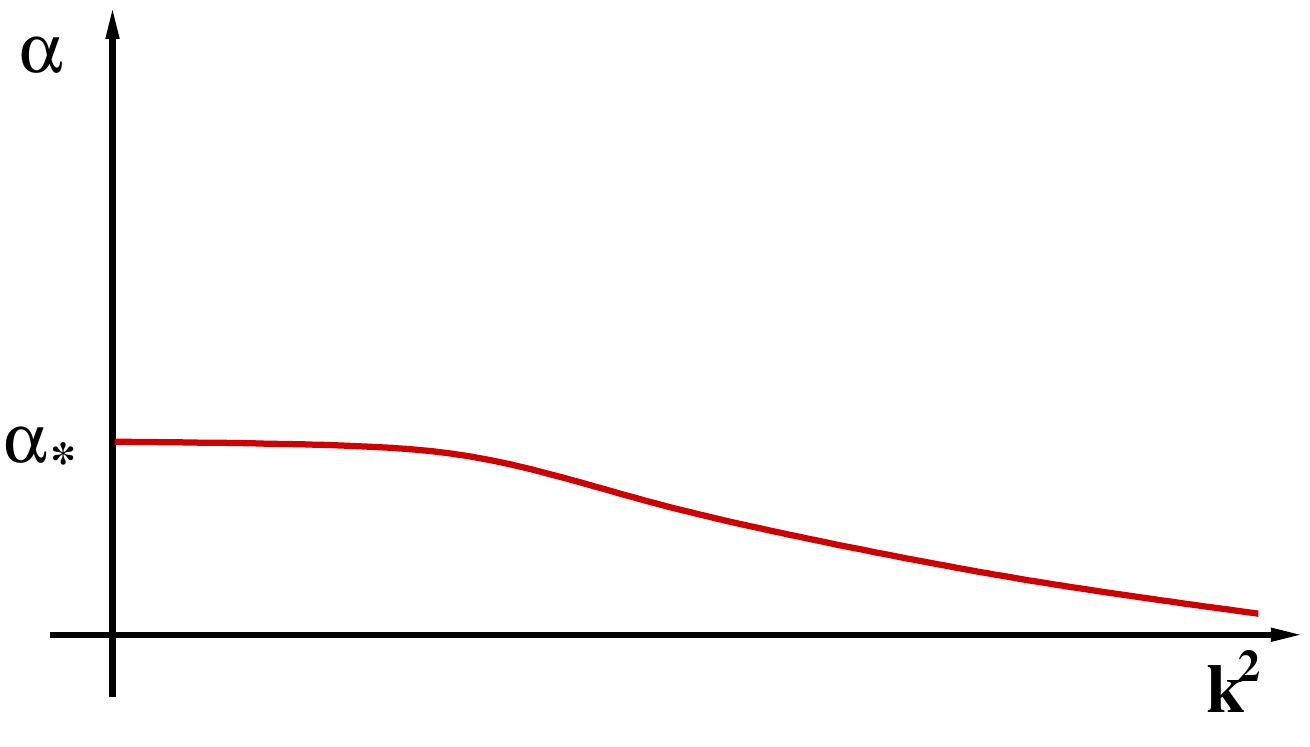}
  \caption{In asymptotically free IR-conformal field theories, for example in the Banks-Zaks limit \cite{Banks:1981nn}, the coupling constant runs at high energies, but reaches a fixed point $\alpha_*$  at low energies. }
  \label{f3}
  \end{figure}

{\em Total disappearance of IR renormalons}: In SQCD in the Seiberg conformal window \cite{qd}
the infrared limit of the theory is conformal. Moreover, near the  edges of the window either the original theory or
its dual is weakly coupled. Since the gauge coupling ceases running in the infrared (Fig. \ref{f3}), the factorial divergence associated with 
renormalons is cut off. This is easy to see from the renormalon generating integral
   \begin{eqnarray}
 \int_{0}^{Q^2}    dk^2\frac{k^2\alpha_s(k^2)}{(1+ k^2/Q^2)^3}
   \end{eqnarray} 
if $\alpha_s(k^2) \to \alpha_{s*}$ at small $k^2$. One can integrate all the way to zero momentum with no singularity on the way, thus, the integration is ambiguity free, and the $\alpha_s(Q^2)$ series is not factorially divergent. 
Hence, there is no need for vacuum condensates  to conspire with  IR renormalons.
And indeed, such theories have no localized asymptotic states (particles) and no vacuum condensates.
(This does not mean that all nonperturbative effects vanish. Instantons and IA pairs show up in the color-singlet correlation functions.)

The above statement has, in fact, broader validity.
Consider a  QCD-like gauge theory (non-supersymmetric) with  fermions in the representation ${\cal R}$, possibly  reducible. One can always choose ${\cal R}$ such that the $\beta$ function of the theory has a zero at a small value of $\alpha_s$ \cite{Banks:1981nn}. Then
 the infrared physics is  conformally invariant,  as seen in perturbation theory. 
  This is the well-known Banks-Zaks (BZ) limit \cite{Banks:1981nn}.   
For example, take the number of fundamental flavors close to the asymptotic freedom boundary,
${ N_f}/{N} =  {11}/{2} - \varepsilon$, with small $\varepsilon$. This limit will correspond to a perturbative IR fixed point at $\alpha_s= \varepsilon$.  
   
    This is also consistent with 't Hooft's intuition  \cite{thooft} that IR-renormalons in asymptotically free  QCD-like theories are possibly related to confinement. If $N_f^*$ is the lower boundary of conformal window, then, upon changing $N_f < N_f^*$ to  $N_f > N_f^*$, the IR-renormalon singularities in the Borel plane must completely disappear.    Losing IR-renormalons may be indicative of the confinement-to-conformality transition. 
   
In the lattice formulation of QCD-like theories, stochastic perturbation theory has been used to  show that the leading large-order behavior of perturbation theory is governed by the  gluon condensate \cite{Bali:2014fea}. It would be interesting, if possible, to study large-orders of stochastic perturbation theory in the vicinity of  the boundary of the conformal window, $N_f^*$, to see the (dis)appearance of the condensate, and the associated quantum phase transition. We also note that the convergence of the OPE for unitary conformal field theories has recently been argued using an explicit conformal block decomposition \cite{Pappadopulo:2012jk}.

\vspace{1mm}

{\em Higgsed theories}: In gauge theories in which all gauge bosons are Higgsed, one can maintain the weak coupling regime provided that the Higgsing scale $v\gg\Lambda$, where $\Lambda$ is the dynamical scale.
An instructive example is the Standard model with the Weinberg angle set to zero. The gauge coupling does not run
at small momenta, it is frozen at $v$. Hence, the IR renormalon-induced factorial divergence of the perturbative series does not extend to infinity, and the associated vacuum condenstes must vanish.

\vspace{1mm}

{\em Higher supersymmetries}: Pure $\N=2$  SYM theory (the Seiberg-Witten model)
presents an example of an intermediate situation. Consider, for instance, the $SU(2)$ gauge group. 
At a generic point in the moduli space, $SU(2)$ is Higgsed down to $U(1)$, and there is no renormalon-induced factorial divergence. If $v\gg\Lambda$, identifiable instantons persist, the I-A pairs show up in the K\"ahler metric  \cite{Yung:1996sv} and should be well-defined. 

If $v\sim \Lambda$, the theory is at strong coupling.  
The set of non-vanishing condensates in various correlation functions
is even narrower than in $\N =1$ theories. For instance, gluino condensates of the type (\ref{chiral}) obviously become forbidden by $\N=2$ supersymmetry.

At the monopole (or, equivalently,  dyon) point in the moduli space, the  supermultiplet becomes massless, reviving
the infrared running in the low-energy SQED which remains after Higgsing of the original theory. However, the $U(1)$ theory is infrared free,
while in the UV domain it becomes a part of the original $SU(2)$ theory again. Now, IR renormalons are Borel-summable, while UV renormalons 
(which cannot be cured within SQED itself) become cured upon the UV-completion of the $U(1)$ effective theory
(i.e. its embedding into the full $SU(2)$).

 ${\cal N}=4 $ SYM theory at the origin of the moduli space is conformally invariant. The coupling constant does not run at all and is fixed at its UV value. All 
condensates vanish identically due to the absence of a dynamical scale and particle-like spectrum. 
IR renormalons (as well as UV renormalons) are not expected, and it is plausible that in the  $N=\infty  $ limit there are no singularities in the Borel plane. Then the planar  perturbation theory has a finite radius of convergence and the Borel  transform is an entire function as $N\to\infty$.  As an example, consider the cusp-anomalous dimension. In the large-$N$ limit it has a finite radius of convergence in the 't Hooft coupling $\lambda$,  with a singularity on the negative real axis \cite{Beisert:2006ez}.
%, at $\lambda_c =-1/8$.  
A finite radius of convergence in the $\lambda$ plane implies that the Borel transform of the perturbation theory is an entire function.

\vspace{1mm}

{\em Conclusions}: We have shown that in supersymmetric theories
contrasting  the OPE with the renormalon-induced factorial divergence leads
to non-trivial predictions. In supersymmetric theories VEVs of many operators are forbiddden; the higher
the degree of supersymmetry, the wider is the class of forbidden operators. This fact implies the disappearance of 
certain (quite conventional) renormalon-induced factorial divergences in perturbation series. 
Conversely, the absence of renormalons may signify that certain operators are absent in the OPE, or are well-defined by themselves. This analysis also applies to orbifold/orientifold daughter theories,
to fully Higgsed gauge theories and to those theories (both supersymmetric and not)
which have an IR conformal regime. 
\bigskip

{\em Acknowledgments}: We thank G. Basar, A. Cherman and A. Vainshtein for discussions and correspondence. GD and MS acknowledge support from DOE grants DE-FG02-13ER41989 and 
DE-SC-0011842, respectively.

%\end{document}

\end{document}